# THE LATE EMISSION OF THERMONUCLEAR SUPERNOVAE

PILAR RUIZ–LAPUENTE

*Department of Astronomy, University of Barcelona*
*Martí i Franqués, 1 – E–08028 Barcelona, Spain*
*and*
*Max–Planck–Institut für Astrophysik*
*Karl–Schwarzschild–Str. 1 – D–85740 Garching, Germany*

## 1. Introduction

A number of questions raised during this meeting can be clarified by the study of the emission of SNe Ia at late times. Do WDs explode below the Chandrasekhar mass? Or, on the contrary, are the differences that we see among Type Ia supernovae (SNe Ia) just variations within the paradigm of Chandrasekhar–mass exploding WDs? The evolution of the spectra along the exponential decline of the SN light curve, and the light curve itself at those phases, bring us information on the density profiles, composition gradients, and mixing within the supernova ejecta. Thus, important clues on nucleosynthesis, hydrodynamics, and chemical stratification can be inferred from the study of the late emission. Indeed, the initial steps on this path have already probed the underlying explosion model for thermonuclear supernovae well beyond a mere rough assessment: the studies of Meyerott (1978, 1980) and Axelrod (1980) provided a firm support to the exploding CO WD model for SNe Ia. Late–time studies of the other type of supernovae, the core–collapse ones, have also been crucial for their theoretical understanding. Fransson & Chevalier (1989) discussed the evidence of mixing in Type II supernovae such as SN 1987A from its effect in the late emission. Their analysis of the emission in the explosion model of CO WDs accreting from He stars revealed, as well, that it was inadequate to account for Type Ib supernovae. The model based on massive progenitors was, on the contrary, favored by their calculations. An essential reference to these and other issues on late–time emission is Fransson (1994).

In the present review, we will concentrate on the physical processes which play a role in the formation of the spectrum and light curves of SNe Ia at late phases, and on the utility of those phases to clarify the way in which C+O WD explode. The role of $e^+$, $\gamma$-rays, the paths of the high–energy radiation to thermalization through a number of complex cascades, the final optical display and the information contained in it, makes supernovae at these phases a physical laboratory dealing with a large variety of radiative processes, with energies ranging from MeV down to a few orders of magnitude below one eV. We will show how it is possible to come out with first–order answers, in the way radiation escapes from the labyrinth of plasma processes.

## 2. Late–time approach: how late?

The mere understanding of the spectral evolution of a supernova is a goal by itself. The temporal development of the spectral features and a proper identification of some of them are still a pending task. We can benefit, however, of some simplification if we decide to study a particular phase of the spectral evolution. Approaches to study the photospheric phase are discussed in this volume by various authors. Here, we concentrate on the times when the continuum emission has dropped substantially and nebular emission increases. We will model the SN at those phases; later on they will become the subject of study of our SN remnant colleagues, as the density becomes significantly lower and the radiative shocks originated by the interaction of the ejecta with the interstellar medium play an important role. The nebular phase starts in SNe Ia at about 80 days after explosion. At this age the ejecta become optically thin and a photosphere does not longer exist. This transition proceeds in a continous way and the continuum emission drops while the the intensity of nebular emission from forbidden transitions of Fe–peak elements grows larger. From the condition that the optical depth to continuum processes becomes negligible, $\kappa_c \rho r \ll 1$, and taking into account that the supernova expands homologously with $r = v \times t$, it can be seen that

$$t_{neb} \approx 250 \ \frac{\sqrt{\kappa_c M}}{v_9} \ days \qquad (1)$$

where $\kappa_c$ is the continuum opacity, M is the total mass, and $v_9$ is the velocity in units of $10^9$ cm s $^{-1}$. Thus, about 250–300 days after explosion the spectrum is really nebular.

At this phase the steady state hypothesis for ionization balance and thermal balance hold (see §6). This allows some simplification of the complex physics of nonthermal processes in these non–LTE plasmas. The optical depth for absorption in permitted transitions of the emitted photons is also significantly larger:

$$\tau_{bf} \approx \sigma_{bf} \ n \ r \tag{2}$$

and it is of the order of

$$\tau_{bf} \approx 6.9 \times 10^9 \ \sigma_{UV} \ \frac{M}{(v_9 t)^2} \approx 1 \times 10^5 \ \frac{1}{(v_9)^2} \tag{3}$$

Bound−free transitions will be followed by recombination cascades down to the ground state. Ionization resulting from this recombination plays a significant role.

## 3. The energy source: radioactivity from $^{56}$Co

The radioactive decay

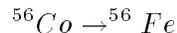

provides the long−lasting luminosity source at this stage. Such a decay gives rise 81% of the time to a $\gamma$−ray photon and 19% to a e$^+$. $\gamma$−ray photons are emitted with a spectrum of energies reaching up to 1.4 MeV. They bear about 96.5 % of the energy of the $^{56}$Co decay. The emitted positrons have an energy spectrum extending up to the endpoint kinetic energy $E_{max}$=1.459 MeV, and they account for 3.5% of the energy of the $^{56}$Co decay.

Compton scattering of the emitted $\gamma$−rays gives rise to a pool of non−thermal electrons called primaries. The optical depth to $\gamma$−rays is given by $\tau_{\gamma} \simeq \sigma_{KN} n_e R$, where $\sigma_{KN}$ is the Klein−Nishina cross section for Compton scattering. Once $\gamma$−ray photons suffer Compton scattering either they do not lose any significant amount of energy (forward scattering), or they lose significantly their energy, becoming readily absorbed (unable to produce further energetic primaries). This has suggested (Sutherland & Wheeler 1984) the adequacy of treating the Compton scattering process as an absorption process, for applications related to the energy deposition of $\gamma$−rays. Swartz, Sutherland, & Harkness (1995) review the result of comparing the purely absorptive approach and the Monte Carlo one. The detailed geometry of an arbitrary 3D distribution of radioactive material can be accounted for within the first approach, saving CPU time as compared with 3D Monte Carlo calculations. However, if one needs to calculate the energy spectrum of the $\gamma$−rays and that of the primaries created by Compton scattering, Monte Carlo calculations are advisable.

The outcome of the calculations is the fraction of radioactive energy deposited in the supernova ejecta:

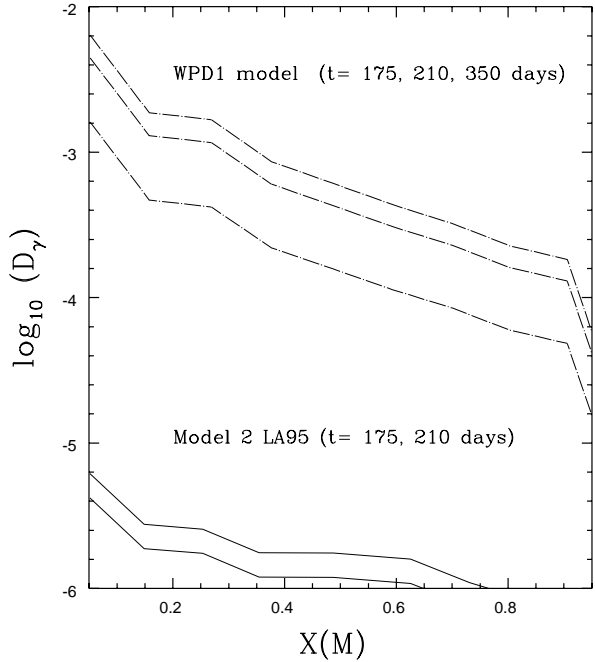

*Figure 1.* Deposition of $\gamma$–rays for two alternative models for subluminous SNe Ia: a sub–Chandrasekhar edge–lit detonation of a 0.7 M$_\odot$ C+O WD (model 2 by Livne & Arnett 1995), and a pulsating delayed detonation model in a Chandrasekhar C+O WD (model WPD1 by Woosley, this volume). The deposition profile is calculated at different times after the explosion

$$\xi(t) = \left(6.76 \times 10^9 \; D_\gamma(t) + 2.72 \times 10^8 \; D_\beta(t)\right) \left(e^{-t/\tau_{Co}} - e^{-t/\tau_{Ni}}\right)$$

$$+ \; 3.91 \times 10^9 \; D_\gamma(t) e^{-t/\tau_{Ni}} \; erg \; g^{-1} \; s^{-1}$$

$$\tag{4}$$

where $\tau_{Ni}$=8.8 days and $\tau_{Co}$=111.26 days are the e–folding times for radioactive decay of Ni and Co respectively, and $D_\gamma$ and $D_\beta$ are the deposition functions of $\gamma$–rays and e$^+$, which are decreasing functions of time. Figure 1 shows $D_\gamma$ for a couple of SNe Ia models.

Positrons start to play a major role in the E input as soon as $\tau_\gamma \ll 0.1$. At these low optical depths (achieved around 250 days after explosion in a 1 M$_\odot$ ejecta), most of the $\gamma$–ray photons escape. An important issue turns then to be the fate of the e$^+$ emitted in the ejecta, and the fraction of escape and deposition of the kinetic energy of those e$^+$. The most energetic e$^+$ will succeed escaping the ejecta. The e$^+$ emitted in the outer layers will also have a higher chance to leave the ejecta without becoming thermalized. The chance for a e$^+$ of a given energy to escape or deposit its kinetic energy in the ejecta depends very much on the configuration of the magnetic field, and requires a numerical evaluation.

The configuration of the magnetic field of the supernova ejecta is therefore a key subject. How is it possible to ascertain which is the most likely configuration of the magnetic field in supernova ejecta? In the first place, one can try to track the final configuration from that of the WD. The WD can have a dipole magnetic field of primordial origin anchored in the inner core, or it can have a dynamo generated magnetic field originated from the presence of charges in motion in the envelope of the WD. How is this magnetic field preserved after the explosion? It is likely that the convective pre–supernova structure will favor a turbulent configuration of the magnetic field, even if it had a dipole configuration in the original WD. From the condition that the kinetic energy density of the ejecta should be much larger than the energy density of the magnetic field (since the kinetic energy density is of the order of the binding energy of the WD), one can estimate the intensity of the magnetic field to be:

$$\frac{B^2}{4\pi\rho v^2} \ll 1 \tag{5}$$

This gives a constraint on the intensity of the magnetic field of $B \ll 10^2 G$ (Chan & Lingenfelter 1993).

The radial configuration of the magnetic field, as suggested by Colgate et al. (1980), would be favored if a dipole magnetic field of the presupernova star were combed out by differential expansion of the supernova ejecta, and the convective pre–supernova structure had not distorted the assumed original dipole configuration.

A empirical way to evaluate the configuration of the magnetic field is to study how much does the late light curve of SNe Ia depart from the full–trapping exponential tail of $^{56}$Co decay. There are reasons to think that a radially combed out magnetic field would give too fast a declining bolometric luminosity as compared with the observations.

## 4. Positrons

The energy spectrum of the e$^+$ has the form:

$$\chi(\epsilon) \propto k(Z, \epsilon)(\epsilon_0 - \epsilon)^2 \epsilon \sqrt{\epsilon^2 - 1} \tag{6}$$

where $\epsilon$ is the (total) positron energy in $m_e c^2$ and

$$\epsilon_0 = E_{max}/m_e c^2 + 1 \tag{7}$$

$$k(Z, \epsilon) = \frac{2\pi\xi}{1 - exp(-2\xi)} \tag{8}$$

with

$$\xi = -\frac{Ze^2}{\hbar v} = -\frac{Z\alpha}{\sqrt{1 - \epsilon^{-2}}} \tag{9}$$

where $Z = 26$, $v$ is the speed of the positron, and $\alpha$ is the fine–structure constant.

The positrons lose their energy in the supernova ejecta mainly by ionization losses. Positrons emitted at a time and place deep in the ejecta where the local density is high will likely lose their kinetic energy and thermalize. If a $e^+$ is emitted at shallow depths or at very late times, such that it cannot completely slow down before leaving the ejecta, it will enter the interstellar medium as a nonthermal particle. If the density is high enough, the thermalized $e^+$ will tend to annihilate giving two 511 keV photons — case the annihilation occurs with a free electron, or they can give either two or three $\gamma$–ray photons —case the annihilation occurs via positronium. The transport and deposition of those $\gamma$-rays is usually treated jointly with that of the $\gamma$-rays emitted in the decay of $^{56}$Co.

The positron energy loss, per unit column depth, X, due ionization of atoms is (Heitler 1954; Blumenthal & Gould 1970; Gould 1972):

$$\frac{dE}{dX} = -\Gamma(E) = -\frac{4\pi r_0^2 m_e c^2 Z}{\beta^2 A m_n}\Pi(E) \tag{10}$$

where $r_0$ is the classical electron radius, $m_n$ is the atomic mass unit, and $Z$ and $A$ are, respectively, the effective nuclear charge and atomic mass of the ejecta material. The $\Pi(E)$ factor is (Berger & Seltzer 1954):

$$\Pi(E) = ln\left(\frac{\sqrt{\gamma - 1}\gamma\beta}{I/m_e c^2}\right) + \frac{1}{2}ln\, 2 - \frac{\beta^2}{12} \times \left[\frac{23}{2} + \frac{7}{(\gamma + 1)} + \frac{5}{(\gamma + 1)^2} + \frac{2}{(\gamma + 1)^3}\right] \tag{11}$$

where $\gamma = 1 + E/m_e c^2$ is the Lorentz factor of the relativistic positron, and $I$ is the effective ionization potential for the ambient atoms in the ejecta. A semiempirical formula for the ionization potential is (Roy & Reed 1968; Segre 1977):

$$I = 9.1 Z \left( 1 + \frac{1.9}{Z^{2/3}} \right) eV \tag{12}$$

Due to the weak dependence of $\Pi(E)$ on $I$, the above approximate formula for I is accurate enough for the practical calculation of the energy loss.

Synchrotron losses by the $e^+$ in the presence of the magnetic field, bremsstrahlung losses, and losses due to Compton scattering off photons are negligible as compared with those arising from ionization/excitation (see Chan & Lingenfelter 1993). Thus, these two last processes are dominant in determining the distance travelled by the $e^+$ before losing its kinetic energy.

One can give some typical values, for positrons of different energies, of their *stopping distance*, $d_e$, due to impact ionization and excitation:

$$d_e \equiv \frac{E}{-dE/dX} \approx \frac{3.36}{\rho} \left( \frac{E}{m_e c^2} \right) \frac{A}{Z} (ln \frac{E}{I_0} + 0.15)^{-1} \ cm \tag{13}$$

TABLE 1. Typical values for the stopping distance and the fraction of the envelope travelled by $e^+$ of various energies

|  | 1 keV | 10 keV | 100 keV | 1 MeV |
|---|---|---|---|---|
| $d_e$ | $9.5 \times 10^{11}$ | $5.5 \times 10^{13}$ | $3.8 \times 10^{15}$ | $3.0 \times 10^{15}$ |
| $\xi$ | $3.6 \times 10^{-5}$ | $2.1 \times 10^{-3}$ | 0.15 | 0.11 |

$d_e$: stopping distance (in cm)
$\xi = d_e / R_{env}$ (300 d)

## 4.1. TRANSPORT AND ESCAPE OF POSITRONS

When the positron mean free path is very small as compared with the characteristic radius of the supernova ejecta, it is possible to evaluate the transport of positrons using the diffusion approximation. This is the case when there is a turbulent magnetic field which confines the trayectories of the positrons along its lines. The case of the radially combed out magnetic field is the opposite: the mean free path of the positron is very large and the diffusion approximation is no longer valid: the transport equation has to be solved. For the case where the positron has *zero diffusion mean free path*, there is a critical time after explosion, $t_c(m_i, \gamma_i)$, such that positrons born at $m_i$ with energy $\epsilon_i$ at times $t_i > t_c$ cannot slow down to thermal energies. These positrons survive in the ejecta as a "fast", nonthermal population.

The time $t_c$ is defined from the energy loss equation:

$$\frac{dE}{\Gamma(E)} = -\rho v dt \ ,$$ (14)

which is:

$$\int_{t_i}^{t} \rho(m_i, t') dt' = -m_e c \int_{\gamma_i}^{\gamma} \frac{\gamma}{\Gamma(\gamma m_e c^2)\sqrt{\gamma^2 - 1}} \ d\gamma$$ (15)

From the above relation and the homology relation $\rho \propto t^{-3}$, one calculates the time, $t_f$, at which a positron, born at time $t_i$, will have lost essentially all of its kinetic energy and become "thermalized" (see Chan & Lingenfelter 1993, for more details):

$$t_f = \left( \frac{1}{t_i^2} - \frac{1}{t_c^2} \right)^{-1/2} \ , \quad \text{for } t_i < t_c$$ (16)

where the critical time after explosion is:

$$t_c(m_i, \gamma_i) = \left[ \frac{8\pi m_e c v_{sn}^2(m_i)}{M} \left( \frac{dv_{sn}}{dm} \right)_{m_i} \times \int_1^{\gamma_i} \frac{\gamma}{\Gamma(\gamma m_e c^2)\sqrt{\gamma^2 - 1}} \ d\gamma \right]^{-1/2}$$ (17)

With the decay rate:

$$R(t) = \frac{1}{(\tau_{Co} - \tau_{Ni})} (e^{-t/\tau_{Co}} - e^{-t/\tau_{Ni}})$$ (18)

the fraction of fast positrons surviving at $m_i$ in the ejecta is:

$$f_f(m_i) = \frac{\tau_{Co}\tau_{Ni}}{\tau_{Ni} - \tau_{Co}} \int_1^{\epsilon_0} d\epsilon_i \chi(\epsilon_i) [\frac{1}{\tau_{Co}} \ exp(-t_c(m_i, \epsilon_i)/\tau_{Ni})$$ (19)

$$- \frac{1}{\tau_{Ni}} exp(-t_c(m_i, \epsilon_i)/\tau_{Co})]$$ (19)

$\chi(\epsilon)$ being the positron spectrum (eq. 6).

In Figure 2 we show the deposition function of the e$^+$ for the turbulent configuration of the magnetic field in different models from 200 to 1000 days after the explosion. Values for the epoch of interest to our calculations are given in Table 2. Chandrasekhar models tend to trap more efficiently the energetic e$^+$, achieving lower escape fractions than sub–Chandrasekhar models.

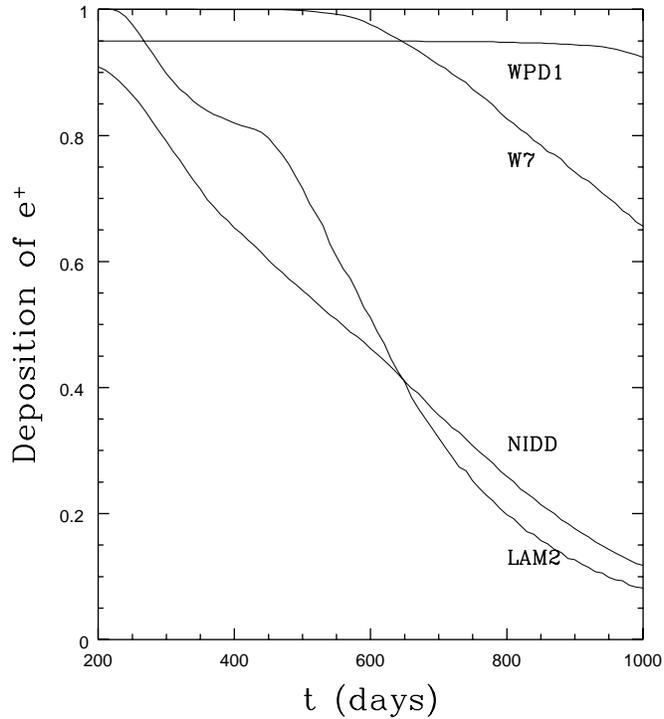

*Figure 2.* Deposition of energy from e$^+$ for various models

TABLE 2. Deposition of e$^+$ for different models[1]

|  | W7 | DD4 | WPD1 | LAM2 | NIDD |
|---|---|---|---|---|---|
| $D_{e+}$ (300$d$) | 100. | 99.9 | 94.9 | 89.7 | 79.1 |
| $D_{e+}$ (350$d$) | 99.9 | 99.9 | 94.9 | 84.6 | 71.5 |

[1] Models are: deflagration model W7 of Nomoto, Thielemann & Yokoi (1984); delayed detonation model DD4 of Woosley & Weaver (1995); pulsating delayed detonation model WPD1 of Woosley (this volume); He–detonation of a 0.7 M$_\odot$ WD (model 2 of Livne & Arnett 1995), and He–detonation of a 1.1 M$_\odot$ WD (Nomoto 1995, here called NIDD)

When the positrons have essentially infinite diffusion mean free paths and they spiral radially through the ejecta, the local fast particle survival fraction has to be determined numerically.

## 5. Nonthermal Processes

The Compton scattering of $\gamma$-rays has created a distribution of energetic $e^-$ called primaries. The energy of the primaries created by Compton scattering off electrons is given by:

$$E_p = \frac{E_\gamma^2 (1 - cos\theta)}{mc^2 + E_\gamma(1 - cos\theta)} \qquad (20)$$

Those primary $e^-$ are able to give rise to impact ionizations, thus producing a cascade of secondary $e^-$. The typical lower energies of the secondaries do not favor further ionization, but they can excite nonthermally the ions in the ejecta or lose their energy through Coulomb scattering off thermal $e^-$.

Treating the energy losses as a continuous process in the case of a plasma dominated by Fe (and Fe–like) ions, one can estimate the ionization rate due to impact ionization by primaries, secondaries, and the excitation rate.

The generation of secondaries is given by:

$$S(E_p, E_s) = \frac{1}{J \; tan^{-1} \left( \frac{E_p - I}{2J} \right) \left( 1 + \left( \frac{E_s}{J} \right)^2 \right)} \qquad (21)$$

where I is the ionization potential and J is an experimentally determined parameter varying with the element (Opal et al. 1971), and somewhat lower than the ionization potential.

The cross sections for the different processes competing in the share of energy are of the form:

$$\sigma_{ion}(E) \propto \frac{ln(E/I)}{E\,I} \qquad (22)$$

for ionization and excitation (Lotz 1967), and

$$\sigma_{cs} \propto \left( \frac{E}{\Delta E} \right) \frac{ln\;\theta}{E^2} \qquad (23)$$

for Coulomb scattering, where $\theta$ depends on $n_e$ and $T_e$.

Using the available cross sections, one can derive the ionization rate by primaries and secondaries. The rates of impact ionization are calculated by evaluating the probability of impacts leading to the final result (ionization of atom/ion l) against all other possible paths through which the primary (or secondary) can lose their energy. For instance, the impact ionization by primaries leading to ions l can be evaluated as:

$$\xi_l^p = n_p \sum_k \int_0^{E_p} \frac{\sigma_{lk}(E)}{L^*(E)} dE \qquad (24)$$

where $\sigma_{lk}$ are the corresponding cross sections for the different transitions (Axelrod 1980; Swartz 1989, 1991), $n_p$ is the rate of production of primaries of energy $E_p$, and $L^*(E)$ is the luminosity loss function which contains the sum of the different channels into which the energy of the primaries can flow: ionization and excitation in all possible transitions, and Coulomb scattering.

Excitation and ionization rates by secondary e$^-$ is calculated after obtaining the energy distribution of secondaries. The rate at which the energy is deposited in heating is evaluated in order to calculate the electron temperature profile.

## 6. The timescales of the radiation processes

The main atomic timescales involved in the emission at these phases are those governing the ionization balance and the thermal balance (since excitation–deexcitation processes occur in a very short timescale, steady state holds). The *recombination timescale* is:

$$\tau_{rec} \sim \frac{1}{n_e \alpha} \qquad (25)$$

where $\alpha$ is the recombination coefficient for a given species and $n_e$ is the electron density.

This can be compared with the *collisional excitation timescale*, given by

$$\tau_{exc} \sim \frac{1}{n_e C_{ik}} \qquad (26)$$

Typically, $\tau_{exc}$ is much smaller than $\tau_{ion}$, and this ensures the possibility of evaluating the statistical equilibrium equations for each ion without the need of coupling all energy levels of five or more ionization stages.

Thermal balance is governed by the dynamical cooling time:

$$\tau_{cool} = \frac{\frac{3}{2} x_e kT}{\alpha} \qquad (27)$$

and the radiative cooling time:

$$\tau_{rad} = \frac{h\nu}{L} \qquad (28)$$

Ionization balance and thermal balance will be in steady state as long as the mentioned timescales will be lower than the evolutionary timescales. These are:

$$\tau_{\,^{56}Co} \sim 111.26 \; days \qquad (29)$$

the *radioactive decay timescale*, and the *expansion timescale*

$$\tau_{exp} = \left| \frac{n}{\dot{n}} \right| = \frac{t}{3} \qquad (30)$$

The *radioactive timescale* reflects the changes in the input energy from $\gamma-$ rays and positrons, and the *expansion timescale* reflects the changes due to homologous expansion and the subsequent decrease in density of the ejecta. At around 200–300 days after the explosion, typical number density $n$ of the ejecta is $10^6$–$10^5$ cm$^{-3}$, and the electron density is of the same order. However, around 500 days $n_e \sim 10^3$ in the central part of the SNIa ejecta and $\tau_{rec} \gtrsim 100$ days for Fe ions. Since this timescale is significantly longer than the evolutionary timescales governing the changes in the nebula (radioactive timescale and expansion timescale), then the steady state hypothesis for ionization processes breaks down and steady state balance is lost. The atomic timescales for recombination processes start to become long at about 500 days after explosion. Those for cooling processes remain short for quite longer times.

## 7. Ionization and rate equations

In a nonthermal nebula, the ionization and recombination processes occur in a longer timescale than excitation and de–excitation, which modify the energy level populations in a shorter timescale. Thus, the ionization balance can be computed first to determine the ion fractions.

The processes to be included to calculate the ionization balance with some degree of accuracy are nonthermal impact ionizations by primaries and secondaries, collisional thermal ionization, and photoionization by emission in recombination cascades.

Since the kinetic energy of the electron is generally much lower than the ionization potential ($kT/Ip \ll 1$), the recombination photons will tend to be emitted very near to the photoionization edge. In this case, and if there were just a single ionic species, a photon emitted by recombination to the ground state of a given ion would be immediately reabsorbed. In the SN ejecta the emitted photon can be reabsorbed in other ionizing transitions of other species. Terms to account for such interactions can be inserted in the same equation of ionization balance. Such equation thus reads:

$$\frac{\partial n_i(t)}{\partial t} = - \left[ \Gamma_i^P + \Gamma_i^S + C_i^T(T) + R_{ik} \right] n_i(t) + \alpha_i(T) n_e n_{i+1}(t) \tag{31}$$

(in steady state balance for ionization, it turns out to be $\partial n_i(t)/\partial t = 0$), where $\Gamma_i^P$ and $\Gamma_i^S$ are the impact ionization rates by primaries and secondaries and

$$R_{ik} n_i = n_i \int_{\nu_0}^{\infty} J_\nu \frac{\sigma_{\nu_{ji}}}{h\nu} d\nu \tag{32}$$

Here, however, the specific intensity of the radiation field $J_\nu$ is only due to the emission of photons originated in recombination cascade decays and the background continuum radiation field is negligible ($J_{\nu(cont)} \simeq 0$).

It is justified to follow the recombination cascade in terms of the possible fates of the photons emitted in those transitions. Photons which are not going to be absorbed on the spot will likely escape further absorptions: reabsorption in permitted transitions has a low optical depth, only enhanced near the resonance region, and reabsorption to ionization has a larger probability to occur near the blue threshold, where the cross section is larger, thus also close to the region where the photon was formed. Here we follow an approximate trapping probability approach which links together all ions and the possibility of recombination giving photons able to ionize other species. It is an escape probability treatment which can account, however, for the effect of recombination to states other than the ground state for all kinds of ions.

The total cross section for a photon $\nu$ to be reabsorbed by an ion is given by:

$$\sigma_{Tj}^{Ph} = \sum_{k=j}^{Z} \frac{n_k}{n} \sigma_{kj}^{Ph}(\nu) \tag{33}$$

and the optical depth for absorption is:

$$\tau_j = \sigma_{Tj}^{Ph} n \Delta r \tag{34}$$

Let $\Phi_{ikj}$ be the probability of a recombination to the a level of ion $j$ giving a photon of a frequency $\nu_k$, able to give rise to ionization of ion $i$. This parameter is well known for the case of recombination to ground state. In general, it is given by:

$$\Phi_{ikj} = \frac{(n_i/n)\sigma_{kj}^{Ph}}{\sigma_{Tk}^{ph}} \Phi_{Tj} \tag{35}$$

where $\sigma_{Tj}^{ph}$ is the total cross section for ionization of any ion susceptible to be ionized by this photon of frequency $\nu_k$, and the function $\Phi$ enters modifying the ionization stage.

The ionization balance reads:

$$\frac{\partial n_i}{\partial t} = -\left[\{\Gamma_i^P + \Gamma_i^S + C_i^T\}n_i + \sum_j \sum_{\nu_k > \nu_i} \Phi_{ikj}\alpha_{j+1}(T)n_e n_{j+1}\right]$$

$$+ (1 - \Phi_{iii})\alpha_{i+1}(T)n_e n_{i+1}$$

$$+ \left[\{\Gamma_{i-1}^P + \Gamma_{i-1}^S + C_{i-1}^T\}n_{i-1} + \sum_j \sum_{\nu_k > \nu_{i-1}} \Phi_{i-1,k,j}\alpha_{j+1}(T)n_e n_{j+1}\right]$$

$$- (1 - \Phi_{i-1,i-1,i-1})\alpha_i(T)n_e n_i = 0 \qquad (36)$$

where $\Gamma_i^P$ is the ionization rate due to primaries, $\Gamma_i^S$ is the ionization rate due to the secondaries, and $C_i$ $(T)$ is the collisional ionization rate.

Such a treatment can be improved by evaluating the flows in the recombination cascades by independent calculations. Monte Carlo evaluation of trapping effects can also help to evaluate the validity of the escape probability approach.

An additional term entering into the ionization balance is the charge transfer term. It is not included here due to the unknown rates of the main charge transfer reactions involving Fe ions. As a general rule, though, charge transfer would tend to modify the fractions of the less abundant ions (the intermediate–mass ions) but produce only a small perturbation in the main components (Fe–peak ions) (Fransson 1994; Swartz 1994).

When evaluating the population of the energy levels, a special attention is required to build up detailed models for the first five ions of Fe. Nonthermal excitation should also be included in the solution of the rate equations: in the He I case, due to the energies of levels above the ground state, nonthermal excitation can play an important role.

## 8. Normal, subluminous, and overluminous SNe Ia

The number of data on SNe Ia studied at these late phases has increased over the last years (Kuchner et al. 1995). The observed spectra of SNe Ia show characteristic forbidden lines of $Fe^{++}$ and $Co^{++}$ evolving in intensity through the late phases. Those observations have shown that the decrease of the emission of the lines at 5800 Å and 6500 Å is consistent with the $^{56}Co \rightarrow {}^{56}Fe$ decay (Kuchner et al. 1995).

There is a diversity among SNe Ia at late phases. As evidenced by the analysis of the spectra, there is a range of velocities over which the forbidden line emission of $Fe^+$ and $Fe^{++}$ extends, among various supernovae. The brightest SNIa, i.e. SN 1991T, has shown the broadest forbidden emission both of Fe and Co ions, whereas the faintest ones show narrow forbidden emission. A range of $^{56}Ni$ masses and electron temperatures has also been derived for different SNe Ia (Ruiz–Lapuente & Filippenko 1993). These studies suggest that the amount of $^{56}Ni$ synthesized in the explosions range from 0.4 to 0.8 $M_\odot$. "Normality" would correspond to about 0.6 $M_\odot$, whereas the overluminous SN 1991T would be located at the top end (0.8 $M_\odot$) and the moderately underluminous SN 1986G at 0.4 $M_\odot$ (Ruiz–Lapuente & Lucy 1992). The very subluminous SN 1991bg can be placed at the lower end of the variation range: this SNIa seems to have synthesized an amount of $^{56}Ni$ as low as 0.1 $M_\odot$ (Ruiz–Lapuente et al. 1993, hereafter RJCF93). The spectra of subluminous SNe Ia indicate differences within the class of subluminous SNe Ia: SN 1986G shows a [Ni II]$\lambda$ 7378 emission, whereas both SN 1991bg and SN 1991F do not show such emission and present a lower ionization stage. SN 1991F can be considered a subluminous SNIa of the 91bg–type (Gómez & López 1995). It is possible that other subluminous SNe Ia like SN 1992bc, SN 1992K, and SN 1971J would have spectra similar to SN 1991bg. Such observations were never performed, however. Thus, the late–time phases of subluminous SNe Ia are poorly documented as compared with those of "normal" SNe Ia, such as SN 1990N, or SN 1992A.

Limits on the fraction of $Fe^0$ and $Fe^{+3}$ have been established respectively from the infrared (Spyromilio et al. 1994) and the UV (Ruiz–Lapuente et al. 1995, hereafter RKPC95) emissions. By studying the nebular emission of SNe Ia it can be concluded that the dominant ionization stage is $Fe^{++}$. The ionization stage is lower in the most subluminous SNe Ia.

The presence of intermediate–mass elements in the Fe–peak region of the ejecta of the faint SN 1991bg suggested that the burning took place at lower densities than in "normal SNe Ia". Detonation models of sub–Chandrasekhar WDs gave a resonable account of the observed spectrum (RJCF93). We will also discuss the nebular spectra of pulsating delayed detonation models as alternative candidates to explain subluminous SNe Ia (§10).

## 9. Diagnostics of the supernova density and mass

Several forbidden transitions of $Fe^+$ and $Fe^{++}$ provide interesting diagnostics of the electron density $n_e$ and electron temperature $T_e$ of the supernova ejecta. We would like to outline here how this can be useful for the present debate between sub–Chandrasekhar and Chandrasekhar models. The emissivities of the lines in a variety of astrophysical situations clearly indicate the physical conditions of the region in which they originate. The approach of infering the electron density profile $n_e(r)$ of the ejecta through the emissivity profile, and extracting reliable information on $n(r)$ is feasable and can carry little error if it is done using the whole information on line emissivities together with the most updated results of the Opacity Project (Pradhan & Berrington 1993; Bautista & Pradhan 1994).

There is a number of emission lines which can be used in the density diagnostics. As in other nebulae, the population of energy levels is going to evolve as the supernova density decreases due to expansion. The upper levels become underpopulated (with departure factors of $10^{-2}$–$10^{-3}$). For some models, this produces some characteristic transitions to show much fainter emission than observed for a given phase. A clear diagnostics of low $n_e$ is associated to the faintness of forbidden emission at $\lambda$ 5200 Å, $\lambda$4300 Å and $\lambda$5000 Å, due to $Fe^+$ $a^4F$–$b^4P$ and $a^4F$–$a^4H$ multiplets. The lower energy terms of these transitions can be significantly depopulated if $n_e$ is low. Interesting line ratios are $\lambda5262/\lambda8617$ and those involving the emission at 4400 Å ([Fe II] $\lambda$ 4416 and other transitions) as compared with the emission in the red and the infrared. Those transitions can help to discriminate the central density of a given supernova. Therefore, this can help to track down the mass. Here we show the evolution of some of these ratios for models with different masses: in Figure 3 we illustrate how the ratio $\lambda5262$ to $\lambda8617$ evolves with time, for different explosion models involving different WD masses, and for two phases: 270 and 300 days after explosion.

There are other transitions of interest for which collision strength calculations are available. The infrared transitions help as indicators of the electron density, in particular the [Fe II]$\lambda$ 8617 to [Fe II]$\lambda$ 1.257 $\mu$m line and, in general, the ratios of lines arising from different energy levels. This approach is very useful, since it overcomes other uncertainties, thus reducing the source of error to the accuracy of collision strengths for the transitions considered. The total mass of the ejecta can be tracked down accurately from the electron density profile, due to the univocal relationship between mass of the exploding WD and central density at any time after explosion.

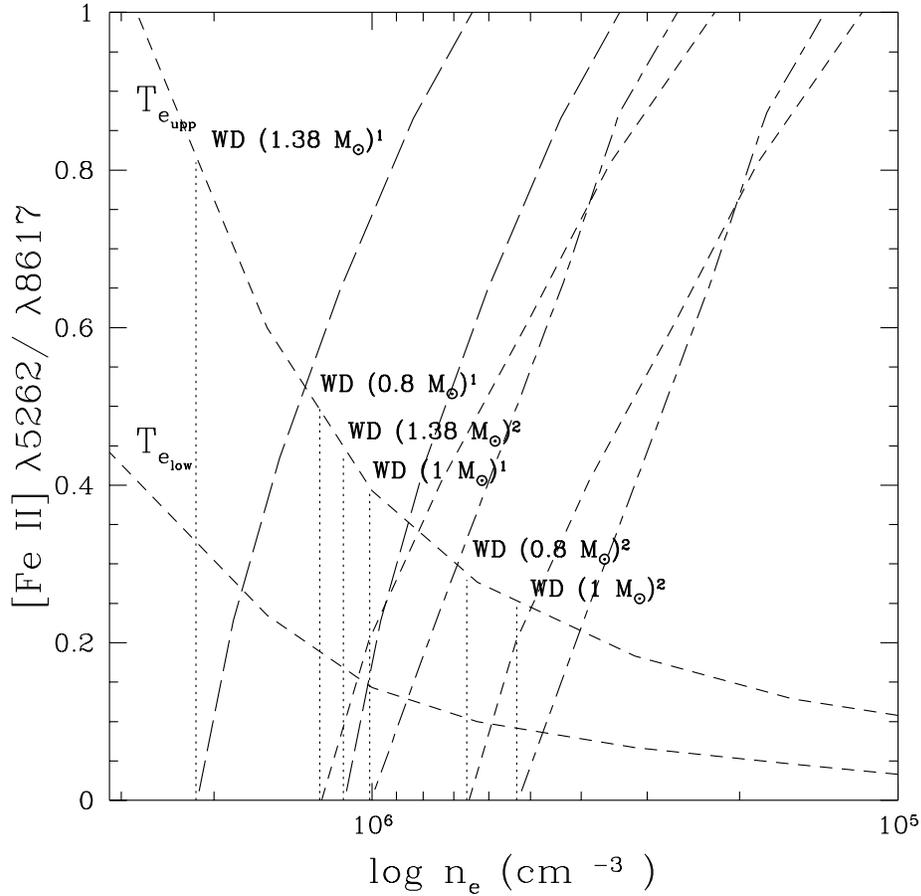

*Figure 3.* Ratio of two characteristic lines for density diagnostics

## 10. Models

Sub–Chandrasekhar models have a number of advantages when the variation of properties among SNe Ia is to be explained. They can account easily for the variation of luminosities and rates of decline of the light curves (Phillips 1993; Riess, Press & Kirshner 1994; Livne & Arnett 1995; Woosley & Weaver 1994). They also explain in an easy way the correlation of these properties (rate of decline of the light curve and maximum brightness) with galaxy type, favoring fainter SNe Ia in early type galaxies as a result of the predominance of older and less massive white dwarfs among the exploding objects (Ruiz–Lapuente, Burkert & Canal 1995).

# SN 1994D in NGC 4526

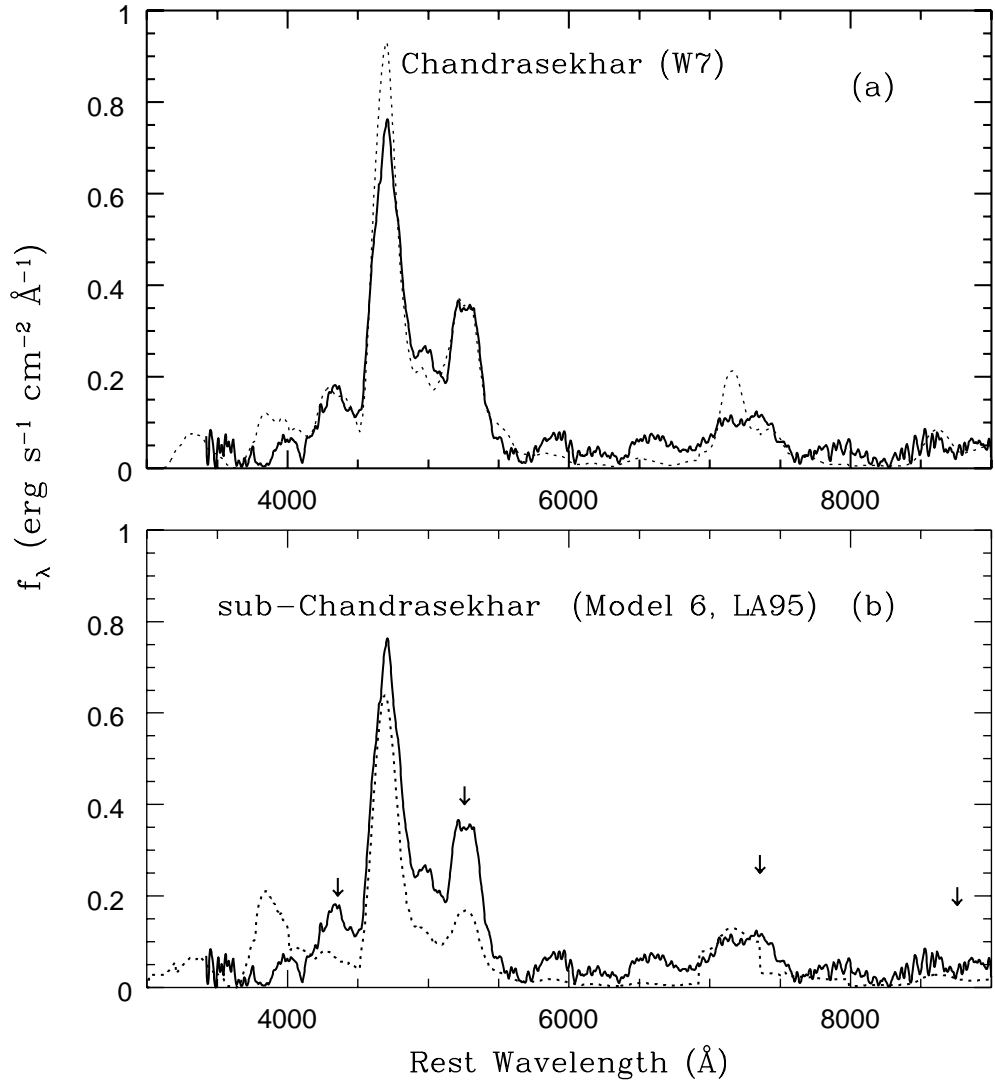

Does, however, the range of sub–Chandrasekhar type of explosions correspond to what is really observed? One would expect, if sub–Chandrasekhar

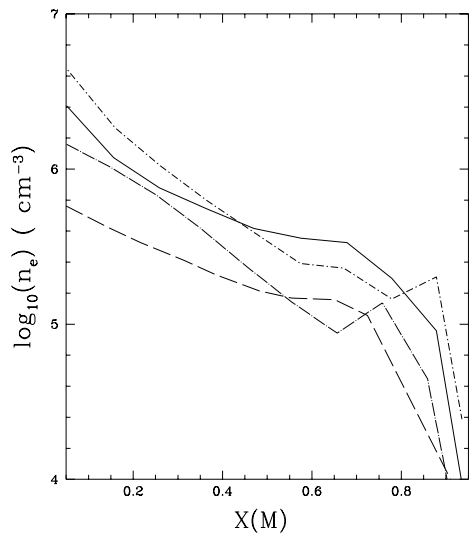

*Figure 5.* Electron density profiles for different models. There is a significant range of central electron densities in models of exploding WDs of various masses at a given phase. This property can help us to discriminate between models

mass WDs exploded in a continuous range from 0.6 to 1.0 $M_\odot$, to have a variety of late–time spectra which has not been seen from the observations so far. Here we judge on the basis of the assessment of 1D He–detonation models (Nomoto 1995, private communication; Woosley & Weaver 1994, and our own calculations) as well as 2D calculations (Livne & Arnett 1995). Both 1D and 2D calculations seem to give similar results.

The central and upper range in mass of the sub–Chandrasekhar C+O explosions give expanding ejecta of density too low as compared with that inferred from observations. The signature of such a low density is the absence of some characteristic emission arising in upper energy levels of $Fe^+$, since those levels become less populated by collisions in lower density ejecta. What happens with the lower mass end of this set of models? The observed spectrum of SN 1991bg and the rate of decline of its light curve seems to be reasonably well explained by the sub–Chandrasekhar models. Both 1D and 2D He detonation models do, however, yield a bit too much $^{56}Ni$ as compared with what is found at late phases in this supernova. The alternative explanation to this supernova, within the Chandrasekhar–mass WD paradigm, can be supplied by pulsating delayed detonation models (see Höflich, Khokhlov & Wheeler 1995, and Woosley, this volume). Model

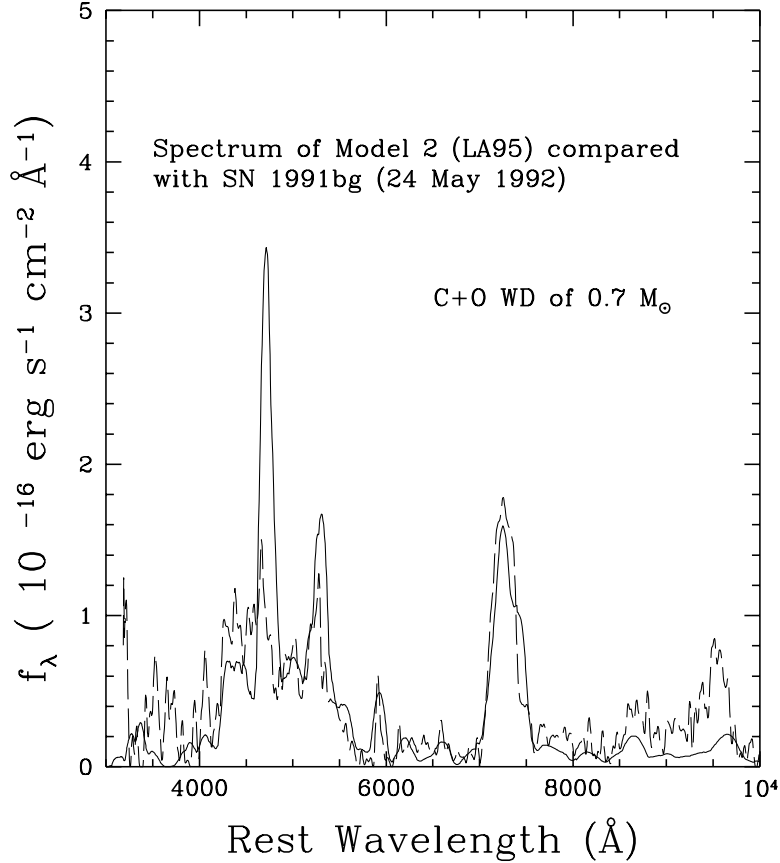

*Figure 6.* A possible model for the subluminous SN Ia SN 1991bg. The solid line shows the model spectrum, and the dashed line the observed one

WPD1 by Woosley (this volume) would be the closest one, among Chandrasekhar models, to SN 1991bg. Ways to discriminate between the sub-Chandrasekhar and Chandrasekhar possibilities are: (a) through the spectral evolution at late phases, i.e. faster in the low WD model than in the pulsating delayed detonation; (b) through the ionization stage shown at these phases, with [Fe III] emission in the low WD explosion model and dominant [Fe II] in the faint WPD1 model; (c) through the presence or not of the [Fe II]$\lambda5557$ emission (and other emissions associated to multiplets $a^4D$–$b^2P$ and $a^4P$–$b^4D$), which become prominent if $n_e$ is about $10^7$ cm$^{-3}$ and Fe$^+$ is the dominant ionization stage, conditions found in Chandrasekhar–mass explosions which synthesized a low amount of $^{56}$Ni (such as some pulsating delayed detonations). Such emission, in contrast, is not expected to be important in a sub–Chandrasekhar envelope, due to the lower $n_e$. Both types of explosions have in common producing faint red SNe Ia at maximum, which are also redder and fainter at late phases than "normal" SNe Ia. A more complete set of observations is needed to ascertain whether 91bg–like events correspond to the lower edge of sub–Chandrasekhar models or to the fainter set of Chandrasekhar pulsating delayed detonations, having in view the above discriminating clues.

## 11. Late light curves

As it has long been argued, the light curves at late phases constitute one more element in the discussion of the total ejected mass in Type Ia SNe (Colgate 1991). At the phase where the thermalization timescale of the radioactive input is short and work to accelerate the envelope is negligible, the luminosity describes the convolution of the exponential decay of $^{56}$Co with the increasing transparency (and escape) of both $\gamma$–rays and $e^+$.

From observations, it is seen that the monochromatic light curves of SNe Ia follow a steeper decline than the full trapping curve of $e^+$ at the times where these dominate the late luminosity (and $\gamma$–rays fully escape). Very few observations allow to reconstruct bolometric light curves: the re-constructed bolometric light curve of SN 1992A (Suntzeff 1995) shows a slight departure below the full trapping curve, whereas in the faint SN Ia SN 1991bg the departure from the full trapping curve (Turatto et al. 1995) seems larger.

Given the nonthermal, NLTE character of the radiating SN ejecta, and the timescales involved in the different processes, the derivation of the monochromatic light curves has to be done through the solution of the spec-tral emission at all wavelength ranges. Here, we show monochromatic light curves up to the phase where time–dependence in ionization and thermal balance starts to play a role. To sample the light curve we have performed spectral calculations every 15–20 days in the tail of the light curve (the colors are given in the B, V, R Johnson system).

Chandrasekhar models for the turbulent configuration of the magnetic field show almost full trapping of the energy of $e^+$ up to 1000 days, whereas edge–lit detonations (sub–Chandrasekhar explosions) predict a departure of the bolometric luminosity below the full trapping line already at 200 days, for the same magnetic field configuration. In the case of a radially combed out magnetic field, both Chandrasekhar and sub–Chandrasekhar models experience an earlier departure from the radioactive tail, which is likely to be too large as compared with most SNe Ia ("normal" ones). Figure 7 shows the result of making the comparison of monochromatic theoretical light curves with observed ones. It is far more illustrative than the comparison of the bolometric light curves, since the monochromatic decays of the light curves inform on how the luminosity in the different spectral bands is distributed and this is much more directly linked to the "goodness" of the underlying model. The agreement of the model in velocity, composition and density space can be, however, only ascertained through the full spectral comparison.

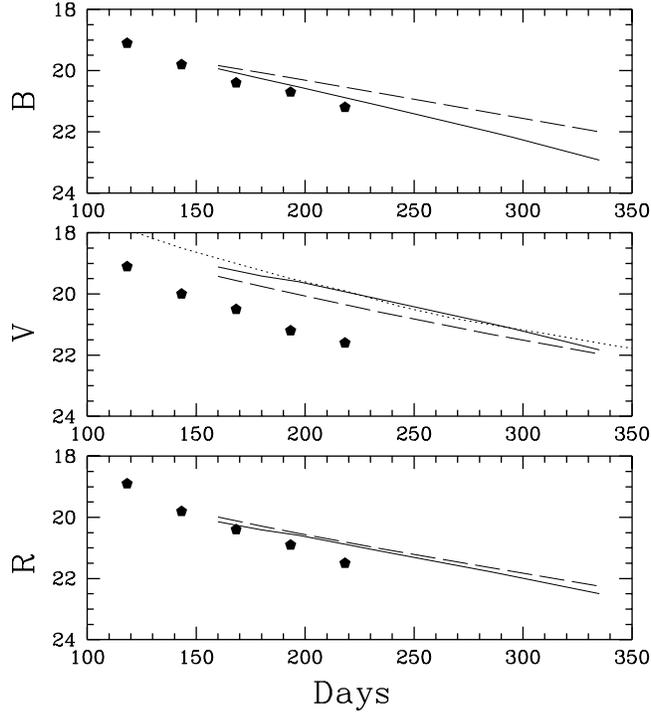

*Figure 7.* The solid lines show the late B V R light curves of model WPD1 (Woosley, this volume), the dashed lines display the same color light curves for model 2 of Livne & Arnett (1995). The solid points correspond to the observations of SN 1991bg (Turatto et al. 1995). The doted line in the central panel is the visual light curve of SN 1992A.

## 12. Final remarks

There is a large number of diagnostics that we can use to obtain information on the validity of a given SN model. Late–time spectra and light curves provide an ideal frame to discuss the mass of the ejecta, kinematics, and nucleosynthesis of thermonuclear SNe. Through the exploration of the ratios of different lines and their dependence on the electron density, we gain insight on whether SNe Ia are sub–Chandrasekhar or Chandrasekhar–mass explosions. *The "present" set of sub–Chandrasekhar explosions seems to give too low density profiles*: about 300 days after explosion the emission coming from upper energy levels is significantly depleted as compared both with more massive models and with the observations. In Chandrasekhar–mass models, emission from excited levels (such as the emission at 4400 Å) is well reproduced. *Red faint SNe Ia at late phases are expected both from low–mass sub–Chandrasekhar explosions and from Chandrasekhar pulsating delayed detonations with little ($\approx$ 0.1 $M_\odot$) $^{56}Ni$. Electron density diagnostics can help to discriminate between these two possibilities.*

A related way to pin down the correct thermonuclear supernova model is through the study of the light curve tails. *Chandrasekhar models trap significantly the $e^+$* and will give bolometric light curve declines close to the full trapping line drawn by the exponential decay of $^{56}$Co. The bolometric light curve of sub–Chandrasekhar models tends to fall below the full–trapping line of $^{56}$Co decay. Monochromatic light curves (B, V, & R) can experience steep declines in some SNe Ia models, as for instance in model 2 of Livne & Arnett (1995), or in the $^{56}$Ni–poor pulsating delayed detonations (such as WPD1 of Woosley, this volume). If clumpiness is present in the supernova ejecta at this phase, escape will be enhanced and the outcome can also be light curves declining faster at late phases. In pulsating delayed detonation models of weak explosions, the formation of dust grains (due to lower temperature), can lead to additional departures in the monochromatic light curves from those for full–trapping. If there is mixing between the region where intermediate–mass elements are present together with Fe (from the previous $^{56}$Ni ) with the O region, the formation of molecules such as $MgSiO_3$ or $Fe_3O_4$ would be favored in models such as WPD1 (the electron temperature is below 2000 K at about 300 days after explosion in those regions).

A more systematic observational coverage of late–time spectra of thermonuclear SNe Ia should be done. So far, the sample of "normal" SNe Ia has been well tracked into the late phases. However, late–time observations of the subluminous SNe Ia have been very unfrequent. It is important to realize that a good nebular tracking of those supernovae will help to discriminate whether WDs below the Chandrasekhar mass do explode or not.

I would like to thank especially Eli Livne and Stan Woosley for providing me with their hydrodynamic models, and for related discussions. My thanks go as well to Leon B. Lucy and Anil Pradhan for informations on available atomic data and discussions on radiation processes in SN nebulae.